\documentclass[12pt]{article}
\usepackage{graphicx}
\usepackage{amsfonts,amsmath}
\usepackage[mathscr]{eucal}
\usepackage{amssymb}
\usepackage{amsthm}
\usepackage{bbold}
\theoremstyle{plain}

\newcommand{\be}{\begin{eqnarray}}
\newcommand{\ee}{\end{eqnarray}}
\newcommand{\bc}{\begin{center}}
\newcommand{\ec}{\end{center}}
\newcommand{\nn}{\nonumber \\}
\newcommand{\lb}{\label}
\newcommand{\p}[1]{(\ref{#1})}

\begin{document}

\begin{titlepage}

\vspace*{0.2cm}

\renewcommand{\thefootnote}{\star}
\begin{center}

{\LARGE\bf  Weak supersymmetric $su(N|1)$ quantum systems}

\vspace{2cm}

{\Large Andrei Smilga} \\

\vspace{0.5cm}

{\it SUBATECH, Universit\'e de
Nantes,  4 rue Alfred Kastler, BP 20722, Nantes  44307, France. }

\end{center}
\vspace{0.2cm} \vskip 0.6truecm \nopagebreak

   \begin{abstract}
\noindent  
We present several examples of supersymmetric quantum mechanical systems with weak superalgebra $su(N|1)$. One of them is the weak $su(N|1)$ oscillator. It has a singlet ground state, $N +1$ degenerate states at the first excited level, etc. Starting from the level $k = N+1$, the system has complete supersymmetric multiplets  at each level involving $2^N$ degenerate states. Due to the fact that the supermultiplets are not complete for $k \leq N$, the Witten index represents a nontrivial function of $\beta$.

This system can be deformed with keeping the algebra intact. The index is invariant under such deformation. The deformed system is not exactly solved, but the invariance of the index implies that the energies of the states at the first $N$ levels of the spectrum are not shifted, and we are dealing with a quasi-exactly solvable system.

Another system represents a weak generalisation of  the superconformal mechanics with $N$ complex supercharges. Also in this case,  starting from a certain energy, the spectrum involves only complete supersymmetric $2^N$-plets. (There also exist normalizable states with lower energies, but they  do not have normalizable superpartners. To keep supersymmetry, we have to eliminate these states.)

   \end{abstract}

\end{titlepage}

\setcounter{footnote}{0}

\setcounter{equation}0

\section{Introduction} 
In Ref.\cite{weak}, we presented a simple supersymmetric oscillator quantum mechanical model enjoying a nonstandard {\it weak} $N=2$ supersymmetric algebra\footnote{In this paper, $N$ counts the number of {\it complex} supercharges.} representing a central extension of the algebra $su(2|1)$. The quantum supercharges and the Hamiltonian of this system can be presented in the form\footnote{In Ref. \cite{weak}, they were written in terms of the fermion operators
$$ \psi_1 = \frac {\chi_1 - \hat{\bar \chi}^2}{\sqrt{2}}, \quad  \psi_2 = \frac {\hat{\bar \chi}^1 +  \chi_2}{\sqrt{2}}, \quad  \hat{\bar \psi}^1 = \frac {\hat{\bar \chi}^1 -  \chi_2}{\sqrt{2}},
\quad  \hat{\bar\psi}^2 = \frac {\hat{\bar \chi}^2 +  \chi_1}{\sqrt{2}},$$
but the form \p{Q}, \p{H} is simpler and can be easily generalised for the case $\alpha 
= 1,\ldots,N$, which is our goal.}

\be
\lb{Q}
\hat Q_\alpha \ =\ (\hat p-ix) \chi_\alpha, \qquad \hat {\bar Q}^\beta = (\hat p+ix) \hat{\bar \chi}^\beta
 \ee
and 
\be 
\lb{H}
\hat H \ =\ \frac {\hat p^2 + x^2}2 + \hat F - 1,
\ee
where  $\alpha,\beta = 1,2$, $\hat p = -i\partial/\partial x $, $ \hat{\bar \chi}^\beta = \partial/\partial \chi_\beta$ 
 and 
 \be
\lb{Fcharge}
\hat F = \chi_\alpha \hat{\bar \chi}^\alpha
 \ee 
is the fermion charge operator.

The supercharges commute with the hamiltonian, but the anticommutator $\{\hat Q_\alpha, \hat{\bar Q}^\beta\}_+$ is not reduced to the Hamiltonian,  involving four other operators:
 \be
\lb{anticom2}
\{\hat Q_\alpha, \hat{\bar Q}^\beta\}_+ \ =\  (2\hat H - \hat Y)\delta_\alpha{}^\beta +  \hat  Z_\alpha{}^{\,\beta}\,,
 \ee
where $\hat Y = \hat F-1$ and 
\be
\lb{Z2}
\hat Z_\alpha{}^\beta \ =\ 2\chi_\alpha \hat{\bar \chi}^\beta - \delta_\alpha^{\ \beta} \hat F.
\ee
Now, $\{\hat Q_\alpha, \hat Q_\beta\}_+$ vanishes and the nonzero commutators of the algebra are
  \be
\lb{alg2}
[\hat Q_\alpha, \hat Z_\beta{}^\gamma] &=& \delta_\beta{}^\gamma \hat Q_\alpha  - 2\,\delta_\alpha{}^\gamma  \hat Q_\beta, 
\qquad [\hat{\bar Q}^\alpha, \hat Z_\beta{}^\gamma] \ = \ 2\,\delta_\beta{}^\alpha \hat{\bar Q}^\gamma -  \delta_\beta{}^\gamma \hat{\bar Q}^\alpha , \nn
\,[\hat Q_\alpha, \hat Y] &=& - \hat Q_\alpha, \qquad [\hat{\bar Q}^\alpha, \hat Y] \ =\   \hat{\bar Q}^\alpha, \nn
 \,[\hat Z_\alpha{}^\beta, \hat Z_\gamma{}^\delta] &=& 2 \left(\delta_\gamma{}^\beta \,\hat Z_\alpha{}^\delta -
\delta_\alpha{}^\delta \,\hat Z_\gamma{}^\beta \right)\,.
 \ee
The last line in \p{alg2} describes $su(2)$ and $\hat Y$ is an extra $U(1)$ charge.

The algebra thus includes two singlet operators, $\hat H$ and $\hat Y$, and represents a {\it central extension} of the $su(2|1)$ algebra.

It is the same algebra that describes the $N=1$ $4D$ supersymmetric field theories placed on the spatial 3-sphere \cite{Sen}. One can write down and study the related supersymmetric  quantum mechanical systems \cite{Romel2,weak-index}, but the simplest SQM system \p{Q}, \p{H} stays apart being  not related to any field theory. 

The spectrum of the Hamiltonian \p{H} involves a singlet bosonic ground state with energy $E = -1/2$ (if the anticommutator $\{\hat Q_\alpha, \hat{\bar Q}^\beta\}_+$ does not give the Hamiltonian, there is no requirement that the vacuum energy is zero), a bosonic and two fermion states with energy $E = 1/2$ (the fourth component of the supersymmetric multiplet with fermion charge $F=2$ is absent here because the fermion states with $E = 1/2$  are annihilated by the action of the supercharges) and the full supersymmetric quartets of states at the levels $E = 3/2, 5/2,\ldots$ The Witten index acquires in this case the contributions from the ground state {\it and} the first excited states
and reads \cite{weak-index}

\be
\lb{Witind}
I_W \ =\ \left \langle \left \langle e^{-\beta \hat H} \right \rangle  \right \rangle
\ \stackrel{\rm def}= \  {\rm Tr} \left\{ (-1)^F  e^{-\beta \hat H}  \right\} = 2 \sinh (\beta/2),
 \ee
representing a nontrivial function of $\beta$. This index has the same nature as the so-called {\it superconformal} index \cite{Romel1} that has been intensely studied in recent years (see  \cite{Rastelli} for the review).

A new remark is that the system \p{Q}, \p{H} admits a natural generalization when one allows for the index $\alpha$ to run from 1 to $N$ with arbitrary $N$.
We will describe this generalization in the next section.

The system \p{Q}, \p{H} can be continuously deformed to include nontrivial interactions so that the algebra \p{anticom2}, \p{alg2} is left intact. The index \p{Witind} stays invariant under such deformation\footnote{This invariance known as {\it localisation} (see Ref. \cite{Pestun} for a review) is traditionally explained in the path integral language, but in Ref.\cite{weak-index} we presented a very transparent proof of this fact using the Hamiltonian language.}. 
The higher $N$ generalization of \p{Q}, \p{H} can also be continuously deformed with keeping the $su(N|1)$ algebra and the value of the index [given by \p{indN}].  The deformed supercharges involve cubic, quintic etc. fermionic terms.

 In Sect. 3, we  present another nontrivial interactive system enjoying 
the weak $su(N|1)$ supersymmetry. This system includes a dimensionful parameter  and is not conformal by itself, but it is related the superconformal quantum mechanical system that includes $N$ complex supercharges satisfying the ordinary supersymmetry algebra. In that case, to keep supersymmetry, we have to keep in the Hilbert space only complete supermultiplets and the index is equal to zero.

\section{$su(N|1)$ oscillator and its deformation}
\setcounter{equation}0
Consider the same system \p{Q}, \p{H}, but assume now that the index $\alpha$ takes $N$ different values. 

The supercharges still commute with the Hamiltonian and the nonzero (anti)commutators are 
  \be
\lb{algN}
\{\hat Q_\alpha, \hat{\bar Q}^\beta\}_+ &=& 2\left(\hat H - \frac {N-1}N  \hat Y\right) \delta_\alpha{}^\beta +  \hat Z_\alpha{}^\beta\,, \nn
\,[\hat Q_\alpha, \hat Z_\beta{}^\gamma] &=& \frac 2N \delta_\beta{}^\gamma \hat Q_\alpha  - 2\,\delta_\alpha{}^\gamma  \hat Q_\beta, 
\qquad [\hat{\bar Q}^\alpha, \hat Z_\beta{}^\gamma] \ = \ 2\,\delta_\beta{}^\alpha \hat{\bar Q}^\gamma -  \frac 2N \delta_\beta{}^\gamma \hat{\bar Q}^\alpha, \nn
\,[\hat Q_\alpha, \hat Y] &=& - \hat Q_\alpha, \qquad [\hat{\bar Q}^\alpha, \hat Y] \ =\   \hat{\bar Q}^\alpha, \nn
 \,[\hat Z_\alpha{}^\beta, \hat Z_\gamma{}^\delta] &=& 2 \left(\delta_\gamma{}^\beta \,\hat Z_\alpha{}^\delta -
\delta_\alpha{}^\delta \,\hat Z_\gamma{}^\beta \right)\,,
  \ee
where now $\hat Y = \hat F - N/2$, $\hat H = \hat p^2 + x^2)/2 + \hat Y$, and 
\be
\lb{ZN}
\hat Z_\alpha{}^\beta \ =\ 2\chi_\alpha \hat{\bar \chi}^\beta - \frac 2N \delta_\alpha^{\,\beta} \hat F.
\ee

The algebra \p{algN} is  a central extension of $su(N|1)$.
When $N=4$, this algebra may be relevant for $6D$  superconformal field theories placed on $S^5 \times \mathbb{R}$, 
but the SQM systems considered in this paper have no field theory connotations. 

The spectrum of the Hamiltonian now involves:
\begin{enumerate}
\item The ground bosonic state $\Psi_0$ with zero fermion charge and the energy $E = (1-N)/2$. It gives the contribution $\exp\{\beta(N-1)/2\}$ to the index
\item A state $\Psi_1$ of zero fermion charge with energy $E = (3-N)/2$ and $N$ fermion states with $F=1$ that are obtained  from $\Psi_1$ by the action of the supercharges $\hat Q_\alpha$. There are no more states at this level because $\hat Q_\alpha \hat Q_\beta \Psi_1 = 0$.
The multiplet is not complete and these states give the contribution
$(1-N)\exp\{\beta(N-3)/2\}$ to the index. 
\item At the level $E = (3-N)/2$ we have a state $\Psi_2$ with $F=0$, $N$ states $\hat Q_\alpha \Psi_2$ and also $N(N-1)/2$ states $\hat Q_\alpha \hat Q_\beta \Psi_2$. The action of the product  $\hat Q_\alpha \hat Q_\beta \hat Q_\gamma$  on $\Psi_2$ gives zero. If $N > 2$, this multiplet is not complete and gives a nonzero contribution to the index. 
\item The last incomplete multiplet has the energy $E = (N-1)/2$. The multiplets with still larger energies are complete including $2^N$ states. Their contribution to the index vanishes. 
\end{enumerate}
Using the identity
$$ \sum_{j=0}^k (-1)^j C^j_N  \ =\ (-1)^k C^k_{N-1} $$
 for the binomial coefficients, we derive
 \be
\lb{indN}
I_W \ =\ \sum_{k=0}^{N-1} (-1)^k C^k_{N-1} \exp \left\{ \beta \left( \frac {N-1}2 - k \right) \right\} \ =\ [2\sinh (\beta/2)]^{N-1}.
 \ee

\subsection{Deformation}
We will describe here a nontrivial continuous deformation of this system that keeps the algebra \p{algN} intact. According to the theorem proven in Ref. \cite{weak-index} (whose validity also follows from localization arguments), the index stays invariant under such deformation.     

It will be more convenient for us to study first a {\it classical} counterpart of this system. The supercharges, Hamiltonian and other operators depend now on commuting ($p,x$) and anticommuting 
($\chi_\alpha, \bar \chi^\alpha$) phase space variables. 
The algebra of commutators and anticommutators is replaced by the algebra of Poisson brackets. The basic Poisson brackets are
\be
\{p, x\}_P = 1, \qquad \{\chi_\alpha, \bar \chi^\beta\}_P = i\delta_\alpha{}^\beta\,.
 \ee
To keep the same functional form of the algebra (with the identification $[\hat A, \hat B] \Rightarrow -i\{A, B\}_P$ if one of the operators is even and $\{\hat A, \hat B\}_+ \Rightarrow -i\{A, B\}_P$ if both of them are odd), we have to pose
\be
\lb{clasop}
Q_\alpha = (p-ix) \chi_\alpha, \quad \bar Q^\beta = (p+ix) \bar \chi^\beta, \quad H = \frac {p^2 + x^2}2 + \chi_\alpha \bar \chi^\alpha, \nn
Y = \chi_\alpha \bar \chi^\alpha, \qquad Z_\alpha{}^\beta = 2\left( \chi_\alpha \bar \chi^\beta - \frac {\delta_\alpha{}^\beta}N \chi_\gamma \bar \chi^\gamma \right).
\ee

Let us first recall what happens in the $N=2$ case.  We seek the deformed   classical supercharges  in the form\footnote{The function $b(x)$ is 2 times larger than the function $B(x)$ introduced in \cite{weak,weak-index}. This change of convention is associated with a different choice of the fermion variables.}
\be
\lb{Qansatz2}
Q_\alpha \ = \  [p- iV(x)] \chi_\alpha -i b(x) \,\chi_\alpha \chi_\gamma \bar \chi^\gamma, \qquad 
   \bar Q^\beta \ = \ [p+ iV(x)] \bar\chi^\beta + i b(x) \,\bar\chi^\beta \chi_\gamma \bar \chi^\gamma, 
\ee
It is easy to see that $\{Q_\alpha, Q_\beta\}_{P}$ and $\{\bar Q^\alpha, \bar Q^\beta\}_{P}$ still vanish.

Now we 
 require that the Poisson bracket $\{Q_\alpha, \bar Q^\beta\}_{P}$ can still be represented in the form \p{anticom2} with  $Y, Z_\alpha{}^\beta$ given in \p{clasop} but with a deformed Hamiltonian. The pure bosonic term and the 4-fermionic term in the  bracket are proportional to $\delta_\alpha{}^\beta$. They contribute to the deformed Hamiltonian. A nontrivial constraint follows from considering bifermionic terms. They include along with  the structure $\propto \delta_\alpha{}^\beta$ also the structure $\propto \chi_\alpha \bar \chi^\beta$. We require that the coefficient of the latter in  $\{Q_\alpha, \bar Q^\beta\}_{P}$ is equal to $2i$, as in the undeformed case. This gives the condition
 \be
\lb{cond2}
V' - bV  \ =\  1\,.
 \ee 
The deformed Hamiltonian reads
\be
\lb{H2-deformed}
H \ =\ \frac{p^2 + V^2}2 + V' \chi_\alpha \bar \chi^\alpha + \frac {b'} 2 (\chi_\gamma \bar \chi^\gamma)^2\,.
 \ee
The property
\be
\{ Q_\alpha, H \}_{P}  =  \{ \bar Q^\beta, H\}_{P} \ =\ 0
\ee
holds, as it should. Also the Poisson brackets of the supercharges with the fermionic phase space functions $Y$ and $Z_\alpha{}^\beta$ have the same form as in \p{alg2}.

\vspace{1mm}

If $N \geq 3$, the Ansatz \p{Qansatz2} does not work well. Imposing the condition \p{cond2}, we can get rid of extra unkosher bi-fermionic terms in $\{Q_\alpha, \bar Q^\beta\}_{P}$, but the 4-fermionic terms $\propto \chi_\alpha \bar \chi^\beta \chi_\gamma \bar \chi^\gamma$ do not easily cancel. For $N=2$ such terms were in fact proportional to $\delta_\alpha{}^\beta$ and gave rise to 4-fermion contribution in $\delta H$, but it is generically not so anymore when we have three or more different fermion flavours. To cancel these terms, we have to impose the condition
$b' -b^2 = 0$, which fixes
$b(x)\  = \ - 1/x$.
Then \p{cond2}  gives $V' + V/x = 1$ with the solution
\be
\lb{V-sol}
V(x) \ =\ \frac x2 + \frac Cx\,.
 \ee
The corresponding weak $su(N|1)$ systems exist, we will study them in Sect. 3, but \p{V-sol} {\it is} not an infinitesimal deformation of the oscillator model with $V(x) = x$. 

To find such deformations, we need to 
 modify the Ansatz to include higher nonlinear  fermionic terms in the supercharges.\footnote{I am grateful to Bruno Le Floch for this insight.} Consider the case $N=3$. We seek the supercharges in the form
\be
\lb{Qansatz3}
Q_\alpha \ =\ (p - iV) \chi_\alpha -ib(x) \,\chi_\alpha \chi_\gamma \bar \chi^\gamma 
-ic(x)\, \chi_\alpha (\chi_\gamma \bar \chi^\gamma)^2
\ee
The terms  $\propto \chi_\alpha \bar \chi^\beta \chi_\gamma \bar \chi^\gamma$ cancel out in the Poisson bracket if the condition
\be
\lb{cond3}
b' - b^2  - 2Vc \ =\ 0
 \ee 
is fulfilled.  This condition together with the condition \p{cond2} admit the solutions $V(x) = x + a(x)$ with an infinitesimal analytic $a(x)$.

For $N=4$, we also have to add the term $-i d(x) \,\chi_\alpha (\chi_\gamma \bar \chi^\gamma)^3$ in $Q_\alpha$ and, to cancel the unkosher 6-fermion terms, impose the extra condition 
\be
\lb{cond4}
c' - 3bc - 3Vd  \ =\ 0
 \ee 
on top of \p{cond2} and \p{cond3}. Again, an infinitesimal deformation $V(x) = x + a(x)$, with $b(x), c(x)$ and $d(x)$ being expressed via $a(x)$, exists.
This procedure is obviously generalized to an arbitrary $N$.
The algebra \p{algN} is kept intact.

\vspace{1mm}

Let us discuss now what happens in the quantum case.
The classical deformed supercharges and the Hamiltonian have their quantum counterparts. As was observed in \cite{howto}, to keep  supersymmetry at the quantum level, it suffices to order the operators in the quantum supercharges following the Weyl symmetric prescription, in which case the Weyl symbol of the anticommutator $\{\hat Q_\alpha, \hat {\bar Q}^\beta\}_+$ coincides with the {\it Gr\"onewold-Moyal bracket} \cite{Moyal} of the classical supercharges,\footnote{The Gr\"onewold-Moyal bracket of two functions on the phase space including the Grassmann dynamical variables is defined as
$$\{A, B\}_{GM} \ =\ 2 \sin \left[ \frac 12 \left( \frac {\partial^2}{\partial p_i \partial Q_i}  - \frac {\partial^2}{\partial P_i \partial q_i}\right) - \frac i2 \left(\frac{\partial^2}{\partial \psi_\alpha \partial \bar \Psi^\alpha} +  \frac{\partial^2}{\partial \bar \psi^\alpha \partial  \Psi_\alpha} \right) \right]
\left.A(p,q;\psi, \bar \psi) B(P,Q; \Psi, \bar \Psi) \right|_{p=P,q=Q; \psi=\Psi, \bar \psi = \bar \Psi }  
$$
($\hbar =1$).} 
\be
\{\hat Q_\alpha, \hat {\bar Q}^\beta\}_W \ =\ \{Q_\alpha, \bar Q^\beta\}_{GM}\,.
\ee
We require then for the unkosher terms to cancel in $\{Q_\alpha, \bar Q^\beta\}_{GM}$, rather than in $\{Q_\alpha, \bar Q^\beta\}_P$. For $N=2$, these brackets coincide, but for $N \geq 3$, there are terms $\propto bc$ (that contribute in the structure $\chi_\alpha \bar \chi^\beta$) and the terms $\propto c^2$ (that contribute in the structure $\chi_\alpha \bar \chi^\beta \chi_\gamma \bar \chi^\gamma$). The conditions \p{cond2} and \p{cond3} are modified to

\be
V' - bV - \frac {bc}2  \ &=& \ 1, \nn
b'-b^2 -2cV - \frac {c^2}2 \ &=& \ 0\,.
 \ee 
These equations are more complicated that \p{cond2}, \p{cond3}, but they still have a solution for infinitesimal $a(x) = V(x)-x$.

We have thus obtained a large family of weak supersymmetric $su(N|1)$ quantum systems parametrized by an arbitrary function $a(x)$. For not too large $a(x)$ (large deformations will be considered in the next section), all these systems have the same Witten index \p{indN}. And that means that all the states at the first $N$ levels belonging to uncomplete supermultiplets  (the total number of such states is $N 2^{N-1}$ ) keep the same energies, from $E_0 = (1-N)/2$ to $E_N = (N-1)/2$, as the oscillator states in the undeformed system. In other words, our systems belong to the large class of {\it quasi-exactly solvable} systems \cite{quasi}, where one can determine exactly the energies of some lowest states in the spectrum, but not of all the states.
 
\subsection{CFIV index}

Besides the Witten index, one can also consider another index introduced in \cite{CFIV} in the context of $2D$  supersymmetric theories,
 \be
\lb{CFIV-def}
I_{CFIV} \ =\ \left \langle \left \langle \hat F e^{-\beta \hat H} \right \rangle \right \rangle.
\ee
This index is invariant under continuous deformations of K\"ahler potential in these theories (though it may depend on the $F$-terms associated with the superpotential).

In our case, the CFIV index can also be introduced and can be shown 
to be  invariant under deformations preserving the algebra \p{anticom2}, \p{alg2}. Indeed, we proved in \cite{weak-index} that not only the Witten index $\langle \langle e^{-\beta \hat H} 
\rangle \rangle$ is invariant under such deformations, but it is true also for a generalized index  
$\langle \langle \hat M e^{-\beta \hat H} 
\rangle \rangle$, provided the operator $\hat M$ commutes with the Hamiltonian and at least one pair of the supercharges.
An example of such an operator is $\hat M_1 = \chi_1 \hat {\bar \chi}^1$, which commutes with $\hat H$, $\hat Q_{\alpha > 1}$ and 
$\hat {\bar Q}^{\alpha > 1}$. (As we see, the CFIV index only exists for {\it extended} strong or weak supersymmetric quantum mechanics, not for the minimal one.) Then $I_1 = \langle \langle \hat M_1 e^{-\beta \hat H} 
\rangle \rangle$ is invariant under deformations and the same concerns $I_{2,\ldots,N}$ and $I_{CFIV} = \sum_{\alpha =1}^N I_\alpha$.

The complete supermultiplets at the higher excited levels do not contribute  to $I_{CFIV}$ due to the relation\footnote{Cf. a similar statement in \cite{Monin} where $2D$ models in a finite spatial box were considered.}
\be
\lb{CFIV-mult}
I_{CFIV}(\mbox{for a $2^N$-plet}) \ =\  \sum_{F=0}^N  F (-1)^F C^F_N \ = \ 0
\ee
The ground state with $F=0$ also does not contribute, and only the first $N-1$ excited levels (which are not shifted under deformation) do. For example, for $N=2$, only two fermion states with the energy $\beta/2$ contribute, and we derive
\be
\lb{CFIV2}
I_{CFIV}^{N=2} \ =\ -2 e^{-\beta/2}\,.
 \ee
For an arbitrary $N$, the calculation gives
\be
\lb{CFIVN}
I_{CFIV} \ =\ -N e^{-\beta/2} \left(2 \sinh \frac \beta 2 \right)^{N-2}\,.
 \ee

\section{Quasi-superconformal $su(N|1)$ systems}
\setcounter{equation}0
As was mentioned, to fulfil the algebra \p{anticom2}, \p{alg2} for any $N$, one may stay with the Ansatz \p{Qansatz2}, choosing
\be
\lb{cond-quasi}
V(x) \ =\ \frac x2 + \frac C x, \qquad b(x) \ =\ - \frac 1x. 
\ee
Then the quantum supercharges and the Hamiltonian (with which the supercharges commute) take the form 
\be
\lb{Q-weak}
\hat Q_\alpha \ &=& \ \chi_\alpha \left(\hat p - \frac {ix}2  - \frac {iC}x\right)  + \frac ix \,\hat{\bar \chi}^\gamma \chi_\alpha \chi_\gamma, \nn
\hat{\bar Q}^\beta \ &=& \  \hat{\bar \chi}^\beta  \left(\hat p + \frac {ix}2  + \frac {iC}x \right) - \frac ix\, \hat{\bar \chi}^\gamma \hat{\bar \chi}^\beta \chi_\gamma\,. 
 \ee
\be
\lb{H-weak}
\hat H \ =\ \frac {\hat p^2}2 + \frac {x^2}8 + \frac {(N-\hat F+C)(N-\hat F-1+C)}{2x^2} + \frac {\hat F+C-1/2}2 \,,
\ee
with $\hat F$ defined in \p{Fcharge}. Note our choice of the ordering prescription in the cubic term. It is not the Weyl choice, but with this particular choice the value $C=0$ gives, as we will see a bit later, a benign exactly solvable Hamiltonian.

The structure $\propto 1/x$ in the supercharges and the structure $\propto 1/x^2$ in the Hamiltonian bring to mind a {\it superconformal} quantum mechanics \cite{superconf}. Our model is not superconformal, the Hamiltonian involves also the oscillator potential, and the frequency of the oscillator [normalized to $\omega = 1/2$ in Eqs. \p{H-weak} and \p{Q-weak}] brings about a dimensionful parameter, but one can suppress the oscillator term to obtain the model found in \cite{Ivanov}:
\be
\lb{Q-strong}
\hat Q_\alpha \ &=& \ \chi_\alpha \left(\hat p - \frac {iC}x \right) + \frac ix \,\hat{\bar \chi}^\gamma \chi_\alpha \chi_\gamma, \nn
\hat{\bar Q}^\beta \ &=& \ \hat{\bar \chi}^\beta \left(\hat p + \frac {iC}x \right)  - \frac ix \, \hat{\bar \chi}^\gamma \hat{\bar \chi}^\beta \chi_\gamma\,. 
 \ee
and 
\be
\lb{H-strong}
\hat H \ =\ \frac {\hat p^2}2 + \frac  1 {2x^2} (N-\hat F+C)(N-\hat F-1+C) \,.
 \ee
That model enjoys the ``strong" supersymmetry with $N$ complex supercharges satisfying the ordinary algebra $\{\hat Q_\alpha, \bar \hat Q^\beta\}_+ = 2\delta_\alpha{}^\beta \hat H$. Our model \p{Q-weak}, \p{H-weak} represents a weak $su(N|1)$ relative of the IKL model.

Note that the very  existence of  interacting SQM models with an arbitrary number of supercharges is a nontrivial fact. For field theories, this would not be possible.

The Hamiltonian \p{H-strong} describes an unbounded motion with a continuous spectrum. The notion of the index is not defined there.\footnote{In fact, for the superconformal models, the Hamiltonian is not a convenient object. Capitalizing on the existence of extra bosonic and fermionic operators in the superconformal group, one may introduce a new Hamiltonian with the discrete spectrum and calculate its index \cite{Fub-Rab}.}  On the other hand, 
the spectrum of \p{H-weak} is discrete and one can try to define the index and to study its expected dependence on $\beta$ and the expected independence on the parameter $C$. However, the physical picture in this case is quite different from what one could expect, basing on the experience of the preceding section, by the following nontrivial reason.

Consider the case $C =0$. In the sectors $F =N$ and $F = N-1$, the term $\propto 1/x^2$ in the Hamiltonian vanishes and the Hamiltonian is reduced to the ordinary oscillator. The ground state in the sector $F=N$ has the energy $E = N/2$. Its wave function is
 \be
\Psi_0^{F=N}  \ =\ e^{-x^2/4} \prod_{\alpha =1}^N \chi_\alpha\,.
\ee
This state is annihilated by $\hat Q_\alpha$, while the action of $\hat{\bar Q}^\beta$ gives a function of fermion charge $F = N-1$ with the spatial dependence $\sim x  e^{-x^2/4}$. Acting on the wave function $ x  e^{-x^2/4}  \prod_{\alpha =1}^{N-1} \chi_\alpha$ by the operator, say,
$$\hat{\bar Q}^1 \ =\ \hat{\bar \chi}^1 \left( -i \frac \partial {\partial x} + \frac {ix}2 \right) - \frac ix \hat{\bar \chi}^N \hat{\bar \chi}^1 \chi_N$$
(the terms $\propto \hat{\bar \chi}^\alpha \hat{\bar \chi}^1 \chi_\alpha$ with $\alpha <N$ do not contribute),
we may observe that the terms $\propto e^{-x^2/4}$ cancel out and the spatial dependence of the wave function in the sector $F = N-2$ is 
$\sim x^2 e^{-x^2/4}$. Acting further  by 
 the products of $\hat{\bar Q}$ on this function and observing the cancellations, one may go all the way down to the sector $F=0$ and reproduce    all  $2^N$ states in the complete supermultiplet of energy $N/2$. These states are normalizable. The same holds for the supermultiplets generated by the action of $\hat{\bar Q}$ on the excited oscillator states in the sector $F=N$.

 The ground states in the sector $F = N-1$ have energy $(N-1)/2$ and the wave functions of the form
   \be
\Psi_0^{F=N-1}  \ =\ e^{-x^2/4} \prod_{\alpha =1}^{N-1} \chi_\alpha\,.
\ee
Now we notice that, when we apply $\hat{\bar Q}^{\beta = 1,\ldots, N-1}$ to this function, the cubic fermion term gives a nonzero contribution $\propto (1/x)e^{-x^2/4}$. The wave function thus obtained is not normalizable and does not belong to the Hilbert space in which we have to work! As a result, the supersymmetry is in fact broken. To keep it, we have to eliminate from the Hilbert space not only non-normalizable states, but also the states that do not have normalizable superpartners \cite{Hilbert}. As a result, the spectrum involves only the complete $2^N$-plets of states with the energies starting from $E = N/2$. 

If $C \neq 0$, the Schr\"odinger equation in the sector $F = N$ reads
\be
\left[ - \frac 12 \frac {\partial^2 }{\partial x^2} + \frac {x^2}8 + \frac {C(C-1)}{2x^2} + \frac {N - 1/2 +C}2 \right] \Psi \ =\ E\Psi \,.
 \ee
One can write two formal solutions to this equation,
\be
\lb{grstN}
\Psi_1^{F=N}(C)   &=& x^C e^{-x^2/4} \prod_{\alpha =1}^N \chi_\alpha  \quad {\rm with\ energy} \quad E = N/2 + 1-C\,, \nn
\Psi_2^{F=N}(C)  &=& x^{1-C} e^{-x^2/4} \prod_{\alpha =1}^N \chi_\alpha \quad {\rm with\ energy} \quad E = N/2+C\,,
\ee
which are normalized when $-1/2 < C < 3/2$. Other eigenstates
cannot be found analytically and the model is in some sense quasiexactly solvable. However, the functions \p{grstN} involve branching points at $x=0$, the Hamiltonian is singular there, and the Sturm-Liouville problem on the line $x \in (-\infty, \infty)$ is not well defined in this case.\footnote{When $C = 0$, it {\it is} well defined in the Hilbert space including the oscillator wave functions in the sector $F=N$ and their superpartners in spite of the singularities of the Hamiltonian in the sectors $F < N-1$. It is so because  these superpartners  behave as $\sim x^2$ or $\sim x^3$ at $x=0$, and $\hat H \Psi$ are not singular.}

 The last comment is the following. Consider the product of all supercharges,
$$\hat S \ =\ \prod_{\alpha=1}^N \hat Q_\alpha\,.$$
This operator is nilpotent. The anticommutator $\hat {\cal H} = \{\hat S, \hat {\bar S}\}$ commutes by construction with $\hat S$ and $\hat{\bar S}$ so that the triple $\hat S, \hat{\bar S}, \hat {\cal H}$ satisfies the standard unextended supersymmetry algebra. The operator $\hat S$ represents a polynomial in momentum, which defines so-called {\it nonlinear supersymmetry} \cite{Andr}, alias {\it $N$-fold supersymmetry} \cite{jap}. In our language, the operators $\hat S$ and $\hat {\bar S}$ relate the sectors $F=0$ and $F=N$. 
 
 It is clear that, keeping in the Hilbert space only these sectors, any extended supersymmetry including weak supersymmetry unravels such nonlinear supersymmetric structure. Inverse is not true, however. There are many systems  with nonlinear supercharges that are not related to any extended  SQM system.

We considered here  only  the simplest $su(N|1)$ models with one real bosonic degree of freedom. Evidently, also $su(N|1)$ models involving more variables should exist and they do. 

One of them, the complex $su(N|1)$ oscillator was suggested in \cite{newSid}. Unfortunately, only the case $N=2$ when this model is equivalent to the ordinary dimensionally reduced free  Wess-Zumino model  enjoying strong supersymmetry, was studied  in detail there. For $N>2$, it is not clear how can one deform the oscillator model to derive a nontrivial weak supersymmeric Hamiltonian. 

Another model, similar to \p{H-weak}  but with larger number of variables, was written in \cite{Armen}. However, the operator that the authors of that paper called ``Hamiltonian" was not actually a proper supersymmetric Hamiltonian. This operator is not related to a supersymmetric Lagrangian by the Legendre transformation and does not commute with the supercharges. Probably, it represents a linear combination of the true Hamiltonian and an extra $U(1)$ charge, so that the true symmetry of the model is $su(N|1)$ with central extension, rather than just $su(N|1)$. Maybe the same concerns the $su(4|1)$ models suggested in \cite{ILS}. Further study of all these interesting and not yet clarified  questions is highly desirable.

    \section*{Acknowledgements}
    
    I am indebted to Sergei Fedoruk, Evgeny Ivanov, Gregory Korchemsky, Bruno Le Floch, Armen Nersessian, Stepan Sidorov, Vyacheslav Spiridonov and Arkady Vainshtein for illuminating discussions  and  valuable remarks. 

I acknowledge the inspiring scientific environment during the program {\it Confinement, flux tubes and large $N$} in KITP, Santa Barbara, where I've started this research. It was supported in part by the National Science Foundation under Grant No. NSF PHY-1748958.

\end{document}